\documentclass[twocolumn,
amssymb, amsmath,superscriptaddress, nobibnotes, aps, dvipdfmx, showpacs]{revtex4-1}

\usepackage[dvipdfmx]{graphicx}  
\usepackage{dcolumn}   
\usepackage{bm}        
\usepackage{amssymb}   
\usepackage{here}
\usepackage{amsmath}
\usepackage{comment}
\usepackage[dvipdfmx]{color}
\usepackage{braket}
\usepackage{mathrsfs}
\usepackage[dvipdfmx]{graphicx}
\usepackage{amsmath}
\usepackage{ulem}
\usepackage{fancybox}
\usepackage{xspace}
\usepackage{amsfonts}
\usepackage{multirow}
\usepackage{amsthm}
\usepackage{setspace}
\usepackage[labelformat=simple]{subcaption}
\usepackage[dvipdfmx]{hyperref}

\begin{document}

\title{Ferromagnetism in the SU($n$) Hubbard model with a nearly flat band}

\author{Kensuke Tamura}
\email{tamura-kensuke265@g.ecc.u-tokyo.ac.jp}
\affiliation{Department of Physics, Graduate School of Science, The University of Tokyo, 7-3-1 Hongo, Tokyo 113-0033, Japan}

\author{Hosho Katsura}
\affiliation{Department of Physics, Graduate School of Science, The University of Tokyo, 7-3-1 Hongo, Tokyo 113-0033, Japan}
\affiliation{Institute for Physics of Intelligence, The University of Tokyo, 7-3-1 Hongo, Tokyo 113-0033, Japan}

%\date{today}%
%%%%%%%%%%%%%%%%%%%%%%%%%%%%%%%%%%%%%%%%%%%%Abstract%%%%%%%%%%%%%%%%%%%%%%%%%%%%%%%%%%%%%%%%%%%%%%%%%%%%
%%%%%%%%%%%%%%%%%%%%%%%%%%%%%%%%%%%%%%%%%%%%%%%%%%%%%%%%%%%%%%%%%%%%%%%%%%%%%%%%%%%%%%%%%%%%%%%%%%%%%%%%
%:abstract
\begin{abstract}
We present rigorous results for the SU($n$) Fermi-Hubbard model on the railroad-trestle lattice.
%a one-dimensional Tasaki lattice. 
We first study the model with a flat band at the bottom of the single-particle spectrum and prove that the ground states exhibit SU($n$) ferromagnetism when the total fermion number is the same as the number of unit cells. 
We then perturb the model by adding extra hopping terms and make the flat band dispersive.
 Under the same filling condition, it is proved that the ground states of the perturbed model remain SU($n$) ferromagnetic when the %lowest 
bottom band is nearly flat. 
 This is the first rigorous example of the ferromagnetism in nonsingular SU($n$) Hubbard models in which both the single-particle density of states and the on-site repulsive interaction are finite.
\end{abstract}
\maketitle

%%%%%%%%%%%%%%%%%%%%%%%%%%%%%%%%%%%%%%%%%%%%%%%%%%%%%%%%%%%%%%%%%%%%%%%%%%%%%%%%%%%%%%%%%%%%%%%%%%%%%%%%
%%%%%%%%%%%%%%%%%%%%%%%%%%%%%%%%%%%%%%%%%%%%Introduction%%%%%%%%%%%%%%%%%%%%%%%%%%%%%%%%%%%%%%%%%%%%%%%%
%:introduction
\section{Introduction} 
\label{sec:intro}
Strongly correlated electron systems, in which the Coulomb interaction between electrons plays an essential role, can exhibit a variety of phenomena such as ferromagnetism, antiferromagnetism, and superconductivity. The Hubbard model has been introduced as a minimum model to describe such systems \cite{Kanamori1963, Gutzwiller1964, Hubbard1963}. 
Dispite its apparent simplicity, the intricate competition between the kinetic and the 
on-site Coulomb terms in the model is hard to deal with analytically. 
%or numerically. 
So far exact/rigorous results have been mostly limited to one dimension~\cite{essler2005one} or systems with special hopping and filling~\cite{Tasaki1992, Mielke1993, Mielke1993a, Tasaki1997, Tasaki2019mb, Derzhko2015}.\par

Recently, it has become possible to simulate the Hubbard model using ultracold atoms in optical lattices~\cite{Kohl2005, Jordens2008, Schneider2008}. 
Furthermore, it was proposed theoretically~\cite{Honerkamp2004} and demonstrated experimentally~\cite{Taie2012} that multi-component fermionic systems with SU($n$) symmetry can be realized in cold-atom setups.
These systems are well described by the SU($n$) Fermi-Hubbard model, in which each atom carries $n$ internal degrees of freedom. When $n=2$, the model reduces to the original Hubbard model with spin-independent interaction. 
%in which case it is referred to as the SU(2) Hubbard model. 
%it was demonstrated that the SU($6$) Hubbard model is feasible with nuclear spin states~\cite{Taie2012}.  
%nuclear spin degrees of freedom  
Although the SU($n$) ($n>2$) symmetry has been less explored in the condensed matter literature, there is a growing interest in recent years in studying the SU($n$) Hubbard model theoretically. 
%because of its experimental realization. 
%Therefore, there is a growing theoretical interest in the SU($n$) Hubbard model because of realizations of the model.
For example, it is argued that the SU($n$) Hubbard model can exhibit exotic 
%novel 
phases that do not appear in the SU($2$) counterpart~\cite{cazalilla2009ultracold, Honerkamp2004, Chung2019}.
Besides, enlarged symmetry other than SU($n$), such as SO(5),  
%symmetryin atomic systems 
in higher-spin systems has also been discussed~\cite{wu2003exact}.
\par
%However, it is also hard to study the SU($n$) Hubbard model as well as the SU($2$) Hubbard model, in spite of theoretical interests in it. 
%However,
The SU($n$) Hubbard model is, in general, harder to theoretically study than the SU(2) Hubbard model.
It has been reported that the Nagaoka ferromagnetism~\cite{Nagaoka1966, Tasaki1989}, which is the first rigorous result for the SU($2$) Hubbard model, can be generalized to the case of SU($n$)~\cite{Katsura2013, Bobrow2018}.
Flat-band ferromagnetism is another example of rigorous results for the SU(2) Hubbard model.
Here, a flat band refers to a structure of single-particle energy spectrum which has a macroscopic degeneracy.
%{\bf
%Flat-band can be seen in a class of lattices constructed based on ``line graph'' method~\cite{mielke1991ferromagnetic} or ``cell construction''~\cite{Tasaki1992}.
A tight-binding model with a flat band can be constructed using standard methods such as the line-graph~\cite{mielke1991ferromagnetic} and the cell constructions~\cite{Tasaki1992}.
%}
In the SU(2) case, it is known that if the system has a flat band at the bottom of the single-particle spectrum and the particle number is the same as the number of unit cells, the ground state of the model exhibits ferromagnetism~\cite{Tasaki1992, Mielke1993, Tasaki1997, Li2004}.
An SU($n$) counterpart of the flat-band ferromagnetism has also been discussed recently~\cite{Liu2019}.\par
%\textbf{, in which the existence of transition between paramagnetism and ferromagnetism was also proved 
%On the other hand, in the SU(2) case, it was shown that the model exhibits ferromagnetism even when the lowest band is dispersive as long as it is nearly flat~\cite{Tasaki1995, Tanaka2018}. However, its generalization to the SU($n$) Hubbard model has not been studied. 

In this paper, 
%we focus on the flat-band ferromagnetism as a rigorous result on the SU(2) Hubbard model and present the generalization of it into SU($n$) Hubbard model. 
we consider the SU($n$) Hubbard model on a one-dimensional (1D) lattice called the railroad-trestle lattice
%1D Tasaki lattice
and derive rigorous results. We first treat the model with a flat band at the bottom and prove that the model exhibits SU($n$) ferromagnetism in its ground states provided that the on-site interaction is repulsive and the total fermion number is the same as the number of unit cells. This is a slight generalization of the result obtained by Liu {\it et al}. in~\cite{Liu2019}, in the sense that our hopping Hamiltonian has one more parameter. 
We then discuss SU($n$) ferromagnetism in a perturbed model obtained by adding extra hopping terms that make the flat band dispersive. We prove that this particular perturbation leaves the SU($n$) ferromagnetic ground states unchanged when the band width of the bottom band is sufficiently narrow. This is our main result and can be thought of as an SU($n$) extension of the previous theorem for the SU($2$) Hubbard model with nearly flat bands~\cite{Tasaki1995}.\par 
The rest of this paper is organized as follows. In Sec.~\ref{sec:thm1}, we introduce the SU($n$) Hubbard model with completely flat band and prove that its ground states exhibit SU($n$) ferromagnetism.
In Sec.~\ref{sec:thm2}, we study a model with nearly flat band and prove that the ground states remain SU($n$) ferromagnetic when the repulsive interaction and the band gap are sufficiently large. 
We present our conclusions in Sec.~\ref{sec:conclusion}.
%to the case of SU($n$). 
%We treat a model with flat-band defined on one-dimensional (1D) lattice called 1D Tasaki lattice and show the generalization of flat-band ferromagnetism on this model.
%In our study, the particle number is fixed to the number of unit cells but the model has general hopping parameter to have flat-band. 
%Furthermore, we discuss the ferromagnetism of the model with nearly flat-band which is not completely flat but has small dispersion. 
%In the latter part of this letter, we present an extension the theorem on the ferromagnetism into the SU($n$) Hubbard model.\par 
%%%%%%%%%%%%%%%%%%%%%%%%%%%%%%%%%%%%%%%%%%%%%Set-up%%%%%%%%%%%%%%%%%%%%%%%%%%%%%%%%%%%%%%%%%%%%%%%%%%%%%
%:setup
\section{Model with completely flat band}~\label{sec:thm1}
\begin{figure}[H]
	\begin{tabular}{c}
		\centering
		\begin{minipage}{0.5 \hsize}
		\subcaption{}\label{fig:deltachain}
			\centering
			\includegraphics[width=\columnwidth]{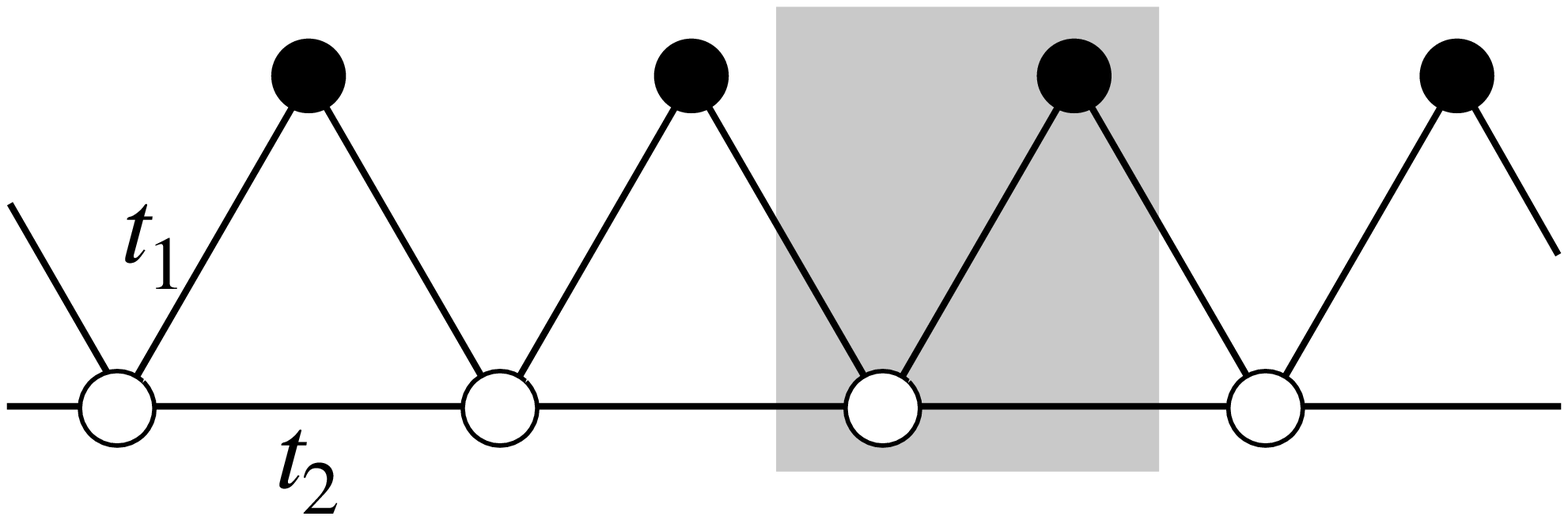}
		\end{minipage}
		
		\begin{minipage}{0.5 \hsize}
		\subcaption{}\label{fig:Energyband}	
			\centering
			\includegraphics[width=\columnwidth]{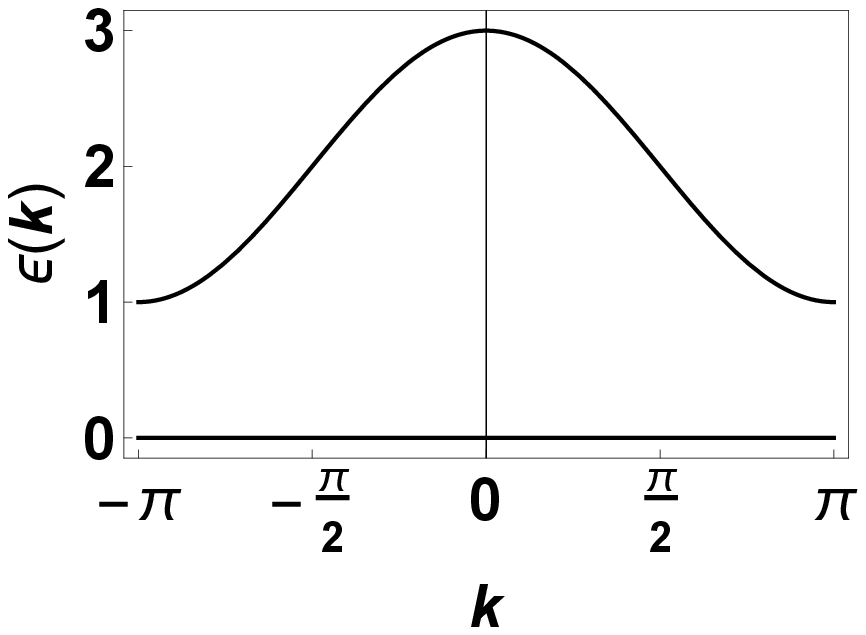}
		\end{minipage}
	\end{tabular}
	\caption{\subref{fig:deltachain} The railroad-trestle lattice with hopping amplitudes $t_{1} = \nu t$ and $t_{2} = \nu^{2} t$. 
	Odd (black) and even (white) sites have on-site potentials $t$ and $2 \nu^{2} t$, respectively.
	The shaded region indicates the unit cell.
	\subref{fig:Energyband} The energy bands for $t = 1, \nu = 1/\sqrt{2}$. The lowest band at zero energy is completely flat. }
	\label{fig:flatband}
\end{figure}

%\subsection{Model 1}
%\label{ssec:model1}
%{\it Model 1.}---
Let $M$ be an arbitrary positive integer and $\Lambda = \{1, 2, \dots , 2M\}$ be a set of $2M$ sites on the railroad-trestle lattice [Fig. \ref{fig:deltachain}].
%Tasaki lattice
We impose periodic boundary condition, so that site $j$ and $j + 2M$ are identified. 
We denote by $\mathcal{E}$ and $\mathcal{O}$ subsets of $\Lambda$ consisting of even sites and odd sites, respectively.
We define creation and annihilation operators $c_{x, \alpha}^{\dag}$ and $c_{x, \alpha}$ for a fermion at site $x \in \Lambda $ with color $\alpha = 1, \dots ,n$.
They satisfy $\{c_{x, \alpha}, c_{y, \beta}^{\dag}\} = \delta_{\alpha, \beta} \delta_{x, y}$.
The number operator of fermion at site $x$ with color $\alpha$ is denoted by $n_{x, \alpha} = c_{x, \alpha}^{\dag} c_{x, \alpha}$.
We consider the SU($n$) Hubbard Hamiltonian
\begin{align}
H_{1} 
&=H_{\mathrm{hop}} + H_{\mathrm{int}} \label{hamiltonian1}, \\ 
H_{\mathrm{hop}}
&= \sum^n_{\alpha=1} \sum_{x,y \in \Lambda} t_{x, y} c_{x, \alpha}^{\dag} c_{y, \alpha}, \\
H_{\mathrm{int}}
&= U \sum_{1 \le \alpha < \beta \le n} \, \sum_{x \in \Lambda} n_{x, \alpha}n_{x, \beta}, \label{hint}
\end{align}
where $t_{2x-1,2x-1} = t$, $t_{2x, 2x} = 2\nu^{2} t$, $t_{2x-1,2x} = t_{2x, 2x-1}= \nu t$, $t_{2x-2,2x}= t_{2x, 2x-2} = \nu^{2} t$, and the remaining elements of $t_{x, y}$ are zero (see Fig. \ref{fig:deltachain}).
The parameters $t, \nu$ and $U$ are positive.
%%%%%%%%%%%%%%%%%%%%%%%%%%%%%%%%%%%%%%%%%%%%%%%FIG%%%%%%%%%%%%%%%%%%%%%%%%%%%%%%%%%%%%%%%%%%%%%%%%%%%%%%
%%%%%%%%%%%%%%%%%%%%%%%%%%%%%%%%%%%%%%%%%%%%%%%%%%%%%%%%%%%%%%%%%%%%%%%%%%%%%%%%%%%%%%%%%%%%%%%%%%%%%%%%
%%%%%%%%%%%%%%%%%%%%%%%%%%%%%%%%%%%%%%%single particle problem%%%%%%%%%%%%%%%%%%%%%%%%%%%%%%%%%%%%%%%%%%
When $U = 0$, the model reduces a tight-binding model and we see that it has two bands with $\epsilon_{1}(k) = 0, \epsilon_{2}(k) = t(2\nu^{2}+1) + 2\nu^{2} t \cos{k}$.
Clearly, the lowest band is dispersionless as shown in Fig. \ref{fig:Energyband}.\par
%%%%%%%%%%%%%%%%%%%%%%%%%%%%%%%%%%%%%definition of spin operators%%%%%%%%%%%%%%%%%%%%%%%%%%%%%%%%%%%%%%%
We define total number operators of fermion with color $\alpha$ and color-raising and lowering operators as $ F^{\alpha, \beta} = \sum_{x \in \Lambda} c_{x, \alpha}^{\dag} c_{x, \beta} $.
Since the Hamiltonian $H_{1}$ has SU($n$) symmetry, they commute with $H_{1}$.
We denote the eigenvalue of $F^{\alpha, \alpha}$ by $N_{\alpha}$.
Since $F^{\alpha, \alpha}$ commute with the Hamiltonian $H_{1}$, the eigenstates of $H_{1}$ are separated into different sectors labeled by $(N_{1}, \dots, N_{n})$.
If the total fermion number $N_{\mathrm{f}} = \sum_{x \in \Lambda} \sum_{\alpha = 1}^{n} n_{x, \alpha}$ is fixed, $N_{\alpha}$ must satisfy $\sum_{\alpha=1}^{n} N_{\alpha} = N_{\mathrm{f}}$. \par
%For each site $x$ in $\Lambda$, 
Now we define a new set of operators 
\begin{align}
&a_{x, \alpha} := -\nu c_{x-1, \alpha} + c_{x, \alpha} -\nu c_{x+1, \alpha} \ \ \ &\text{for} \ x \in \mathcal{E},  \\
&b_{x, \alpha} := \nu c_{x-1, \alpha} + c_{x, \alpha} + \nu c_{x+1, \alpha} \ \ \ &\text{for} \  x \in \mathcal{O}, 
\end{align}
which satisfy 
\begin{align}
\{a_{x, \alpha}, b_{y, \beta}^{\dag}\} &= 0,  \label{ab}\\
\{a_{x, \alpha}, a_{y, \beta}^{\dag}\} &= 
\begin{cases}
\delta_{\alpha, \beta} (\nu^{2} + 2) \ \ &\text{if}\  x  = y , \\
\delta_{\alpha, \beta} \ \nu^{2} \ \ &\text{if}\ x = y \pm 2, \\
0 \ \ &\text{otherwise},
\end{cases} \\
\{b_{x, \alpha}, b_{y, \beta}^{\dag}\} & = 
\begin{cases}
\delta_{\alpha, \beta} (\nu^{2} + 2) \ \ &\text{if}\ x = y, \\
\delta_{\alpha, \beta} \ \nu^{2}\ \ &\text{if}\ x = y \pm 2, \\
0 \ \ &\text{otherwise}.
\end{cases}
\end{align} 
The hopping Hamiltonian $H_{\mathrm{hop}}$ is rewritten in terms of $b_{x, \alpha}$ and $b_{x, \alpha}^{\dag}$ as 
\begin{align}
H_{\mathrm{hop}} = t \sum_{\alpha=1}^{n} \sum_{x \in \mathcal{O}} b_{x,\alpha}^{\dag} b_{x, \alpha} \label{hop1}
\end{align}
and hence positive semi-definite.
The interaction term $H_{\mathrm{int}}$ is also positive semi-definite because $n_{x, \alpha} n_{x, \beta} = \left(c_{x, \alpha} c_{x, \beta}\right)^{\dag} c_{x, \alpha} c_{x, \beta}$.
Therefore, the total Hamiltonian $H_{1} = H_{\mathrm{hop}} + H_{\mathrm{int}}$ is positive semi-definite as well.
From now on, we fix the total fermion number as $N_{\mathrm{f}} = |\mathcal{E}| = M$ and define a fully polarized state as $\ket{\Phi_{\mathrm{all}, \alpha}} := \prod_{x \in \mathcal{E}} a_{x, \alpha}^{\dag} \ket{\Phi_{\mathrm{vac}}}$, where $\ket{\Phi_{\mathrm{vac}}}$ is a vacuum state of $c_{x, \alpha}$.
From the anti-commutation relation (\ref{ab}), we find that $\ket{\Phi_{\mathrm{all}, \alpha}}$ is an eigenstate of $H_{1}$ with eigenvalue zero.
Since $H _{1}\geq 0$, the fully polarized states are ground states of $H_{1}$.
Due to the SU($n$) symmetry, one obtains a general form of degenerate ground states as
\begin{align}
\ket{\Phi_{N_{1}, \dots, N_{n}}} = \left(F^{n, 1}\right)^{N_{n}} \dots \left(F^{2, 1}\right)^{N_{2}} \ket{\Phi_{\mathrm{all},1}}, \label{fully polarized}
\end{align}
where $N_{1} = M -\sum_{\alpha=2}^{n}N_{\alpha}$.
We also refer to states of the form Eq. (\ref{fully polarized}) as fully polarized states~\footnote{
The total number of such states is $\frac{(M+n-1)!}{M! (n-1)!}$.
}.

The first result of this paper is the following: \par
%%%%%%%%%%%%%%%%%%%%%%%%%%%%%%%%%%%%%%%%%%%%%%%%%%%%%%%%%%%%%%%%%%%%%%%%%%%%%%%%%%%%%%%%%%%%%%%%%%%%%%%%
%%%%%%%%%%%%%%%%%%%%%%%%%%%%%%%%%%%%%%%%%%%%%%%%%%%%%%%%%%%%%%%%%%%%%%%%%%%%%%%%%%%%%%%%%%%%%%%%%%%%%%%%
%%%%%%%%%%%%%%%%%%%%%%%%%%%%%%%%%%%%Flat-band ferromagnetism%%%%%%%%%%%%%%%%%%%%%%%%%%%%%%%%%%%%%%%%%%%%
%:Theorem1
%\subsection{Theorem 1.}
%\label{ssec:thm1}
{\it Theorem 1.}---Consider 
%the above Hubbard model 
the Hubbard Hamiltonian (\ref{hamiltonian1}) with the total fermion number $N_{\mathrm{f}} = M$.
For arbitrary  $t>0$ and $U>0$, the ground states of the Hamiltonian (\ref{hamiltonian1}) are the fully polarized states and unique apart from trivial degeneracy due to the SU($n$) symmetry.
%:Proof of thm1

\renewcommand{\proofname}{{\indent \it Proof of Theorem 1}}
\renewcommand{\qedsymbol}{$\blacksquare$}
{\it Proof of Theorem 1.}---Let $\ket{\Phi_{\mathrm{GS}}}$ be an arbitrary ground state of $H_{1}$ with $N_{\mathrm{f}} = M$.
Since the ground state energy is zero, we have $H_{1} \ket{\Phi_{\mathrm{GS}}} =0$.
The inequalities $H_{\mathrm{hop}} \geq 0$ and $H_{\mathrm{int}} \geq 0$ imply that $H_{\mathrm{hop}} \ket{\Phi_{\mathrm{GS}}} =0$ and $H_{\mathrm{int}} \ket{\Phi_{\mathrm{GS}}}=0$, which means that
\begin{align}
&b_{x, \alpha} \ket{\Phi_{\mathrm{GS}}} 
=0\  \ \text{for any $x \in \mathcal{O}$ and $\alpha = 1, \dots n$}, \label{condition1}\\
&c_{x,\alpha} c_{x, \beta} \ket{\Phi_{\mathrm{GS}}} 
= 0\ \ \text{for any $x \in \Lambda$ and $\alpha \neq \beta$}. \label{condition2}
\end{align}
Since $a_{x, \alpha}$ and $b_{x, \alpha}$ obey the anti-commutation relation (\ref{ab}), the condition (\ref{condition1}) implies that $\ket{\Phi_{\mathrm{GS}}}$ does not contain any $b_{x, \alpha}^{\dag}$ operator when it is constructed by acting with creation operators on the vacuum state.
Therefore, it is written as
\begin{align}
&\ket{\Phi_{\mathrm{GS}}} \nonumber \\
& = \! \sum_{\substack{A_{1}, A_{2} ,\dots A_{n} \subset \mathcal{E}\\
\sum_{\alpha =1}^{n} |A_{\alpha}|= M}} \!
f(\{A_{\alpha}\}) \left( \prod_{x \in A_{1}} a_{x,1}^{\dag}\right) \! \!
\dots
\! \left( \prod_{x \in A_{n}} a_{x,n}^{\dag}\right) \! \!
\ket{\Phi_{\mathrm{vac}}},
\end{align}
where $A_{\alpha}$ is a subset of $\mathcal{E}$ and $f(\{A_{\alpha}\})$ is a  certain coefficient.\par
Next, we make use of the condition (\ref{condition2}).
We take an even site $x \in \mathcal{E}$. 
Using the anti-commutation relation $\{c_{x, \alpha}, a_{y,\beta}^{\dag} \} = \delta_{\alpha, \beta} \delta_{x,y}$ and %the condition 
Eq. (\ref{condition2}) we see that 
%$f(A_{1}, \dots , A_{n}) = 0$ 
$f(\{A_{\alpha}\})=0$ if there exist $A_{\alpha}$ and $A_{\beta}$ such that $A_{\alpha} \cap A_{\beta} \neq \emptyset$.
Since $\sum_{\alpha=1}^{n}|A_{\alpha}| = M$ and $A_{\alpha} \cap A_{\beta} = \emptyset$ for $\alpha \neq \beta$, we find that $\cup_{\alpha=1}^{n} A_{\alpha} = \mathcal{E}$.
This means that the ground state is rewritten as 
\begin{align}
\ket{\Phi_{\mathrm{GS}}}
= \sum_{\bm{\alpha}}C(\bm{\alpha}) \left(\prod_{x \in \mathcal{E}} a_{x, \alpha_{x}}^{\dag}\right)
\ket{\Phi_{\mathrm{vac}}}, 
\end{align}
where the sum is over all 
%the 
possible color configurations $\bm{\alpha} = (\alpha_{x})_{x \in \mathcal{E}}$ with $\alpha_{x} = 1, \dots ,n$. 
%on $\mathcal{E}$.
Then we consider the condition (\ref{condition2}) for $x \in \mathcal{O}$.
By using 
\begin{align}
\{c_{x,\alpha}, a_{y, \beta}^{\dag}\} =
\begin{cases}
-\nu \delta_{\alpha, \beta} \ \ &\text{if $y = x\pm1$}, \\
0 \ \ &\text{otherwise},
\end{cases} 
\label{anticom1}
\end{align}
%By using (\ref{anticom1}), 
we get
\begin{align}
&c_{x, \alpha}c_{x, \beta} \ket{\Phi_{\mathrm{GS}}} \nonumber \\
&= \sum_{\substack{\bm{\alpha}\\
	\mathrm{s.t.} \alpha_{p} = \beta, \\
	\alpha_{q} = \alpha}} 
	\nu^{2} \left[C(\bm{\alpha} ) - C(\bm{\alpha}_{p\leftrightarrow q})\right]
\left(\prod_{y \in \mathcal{E}\backslash \{x\pm 1\}} a_{y, \alpha_{y}}^{\dag}\right)\ket{\Phi_{\mathrm{vac}}}, 
\end{align}
where $p=x-1$ and $q = x+1$.
The color configuration $\bm{\alpha}_{p \leftrightarrow q}$ is obtained from $\bm{\alpha}$ by swapping $\alpha_{p}$ and $\alpha_{q}$.
Since all the states in the sum are linearly independent, we find from the condition (\ref{condition2}) that $C(\bm{\alpha}) = C(\bm{\alpha}_{p\leftrightarrow q})$ for all $\bm{\alpha}$ and all $x \in \mathcal{O}$.
As the two localized states on 
%the 
neighboring even sites share an odd site between them, we see that
\begin{align}
C(\bm{\alpha}) = C(\bm{\alpha}_{x \leftrightarrow y}), \label{symmetric}
\end{align}
where $x, y$ are arbitrary different sites in $\mathcal{E}$.\par
To show that 
%such 
states satisfying Eq. (\ref{symmetric}) are the fully polarized states, i.e., SU($n$) ferromagnetic, we introduce a concept of a word~\cite{kitaev2011patterns}. 
A {\it word} $w = (w_{1}, \dots, w_{M})$ is a sequence of integers where $w_{i} \in \{1,\dots,n\}$ for all $i$. 
We denote by $|w|_{\alpha}$ the number of occurences of $\alpha$ in $w$. 
We define the set of words for which $|w|_{\alpha} = N_{\alpha}$ holds as follows: $W(N_{1}, \dots , N_{n}) = \{w | \ |w|_{\alpha} = N_{\alpha},  \ \alpha = 1, \dots , n \}$.
For example, $W(2, 0, 1)$ consists of $(1, 1, 3), (1, 3, 1)$ and $(3, 1, 1)$.
It follows from 
%From the condition 
Eq. (\ref{symmetric}) that the ground state of $H_{1}$ in the sector labeled by $(N_{1}, \dots , N_{n})$ can be written as
\begin{align}
\ket{\widetilde{\Phi}_{N_{1}, \dots , N_{n}}} = \sum_{w \in W(N_{1}, \dots , N_{n})} a_{2, w_{1}}^{\dag} a_{4, w_{2}}^{\dag} \dots a_{2M, w_{M}}^{\dag} \ket{\Phi_{\mathrm{vac}}}.
\end{align}
%Now we show 
%One can show that $\ket{\widetilde{\Phi}_{N_{1}, \dots , N_{n}}} = \ket{\Phi_{N_{1}, \dots, N_{n}}}$ .
Now using commutation relations $[F^{\beta, \alpha}, a_{x, \gamma}^{\dag}] = \delta_{\alpha, \gamma} a_{x, \beta}^{\dag}$ for all $x \in \mathcal{E}$ , we see that 
\begin{align}
\left(F^{2, 1}\right)^{N_{2}} \! \ket{\Phi_{\mathrm{all}, 1}} 
= \! \! \! \! \sum_{w \in W(M-N_{2}, N_{2})} \! \! \!  \! a_{2, w_{1}}^{\dag} a_{4, w_{2}}^{\dag} \dots a_{2M, w_{M}}^{\dag} \ket{\Phi_{\mathrm{vac}}}.
\end{align}
%where $N_{1} = M - N_{2}$. 
By repeating the procedure, we have the desired result $\ket{\widetilde{\Phi}_{N_{1}, \dots , N_{n}}} = \ket{\Phi_{N_{1}, \dots, N_{n}}}$.
%\begin{align}
%&\left(F^{n,1}\right)^{N_{n}} \dots \left(F^{2,1}\right)^{N_{2}} \ket{\Phi_{\mathrm{all}, 1}} \nonumber \\
%&= \sum_{w \in W(N_{1}, N_{2}, \dots, N_{n})} a_{2, w_{1}}^{\dag} a_{4, w_{2}}^{\dag} \dots a_{2M, w_{M}}^{\dag} \ket{\Phi_{\mathrm{vac}}}.
%\end{align}
This proves that the ground states of $H_{1}$ are fully polarized states.
\hspace{\fill}$\blacksquare$
%%%%%%%%%%%%%%%%%%%%%%%%%%%%%%%%%%%%%%%%%%%%%%%%%%%%%%%%%%%%%%%%%%%%%%%%%%%%%%%%%%%%%%%%%%%%%%%%%%%%%%%%%%%%%%%%%%
%%%%%%%%%%%%%%%%%%%%%%%%%%%1D Tasaki lattice with nearly flat-band%%%%%%%%%%%%%%%%%%%%%%%%%%%%%%%%%%%%%
\section{Model with nearly flat band}
\label{sec:thm2}
\begin{figure}[H]
	\begin{tabular}{c}
		\centering
		\begin{minipage}{0.5\hsize}
		\subcaption{}\label{fig:deltachain2}
			\centering
			\includegraphics[width=\columnwidth]{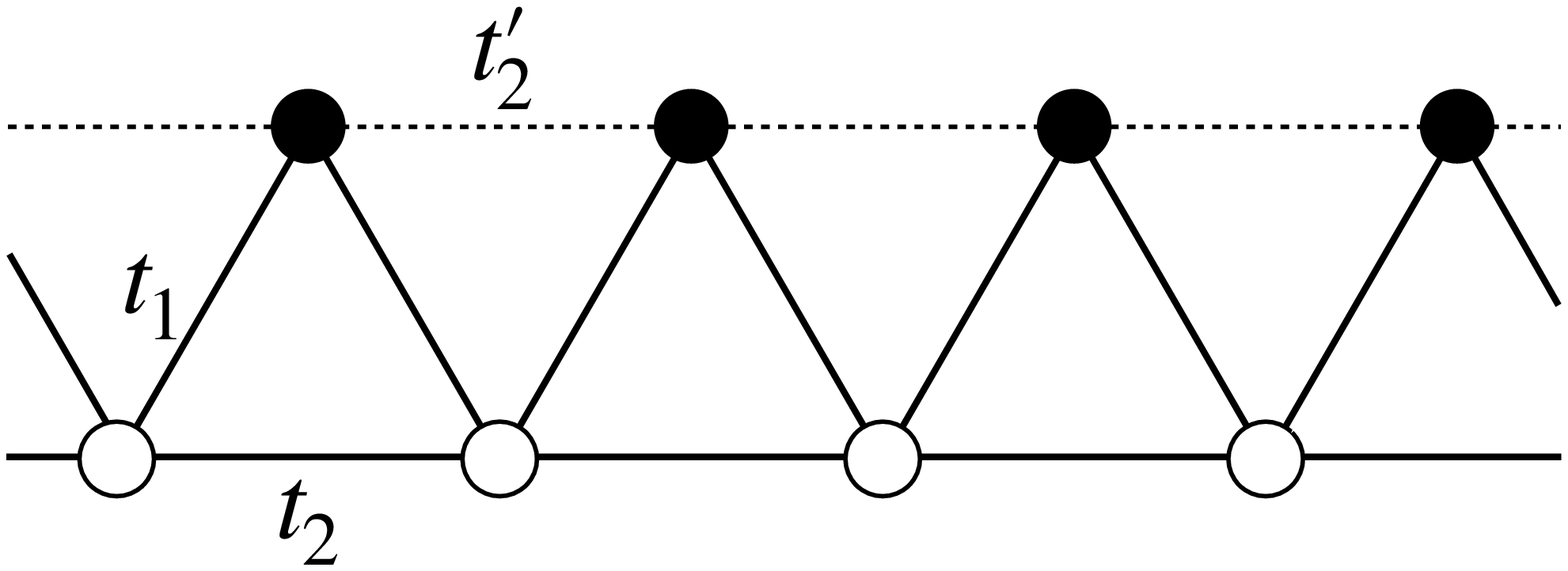}
		\end{minipage}
		\begin{minipage}{0.5\hsize}
		\subcaption{}\label{fig:Energy band2}
			\centering
			\includegraphics[width=\columnwidth]{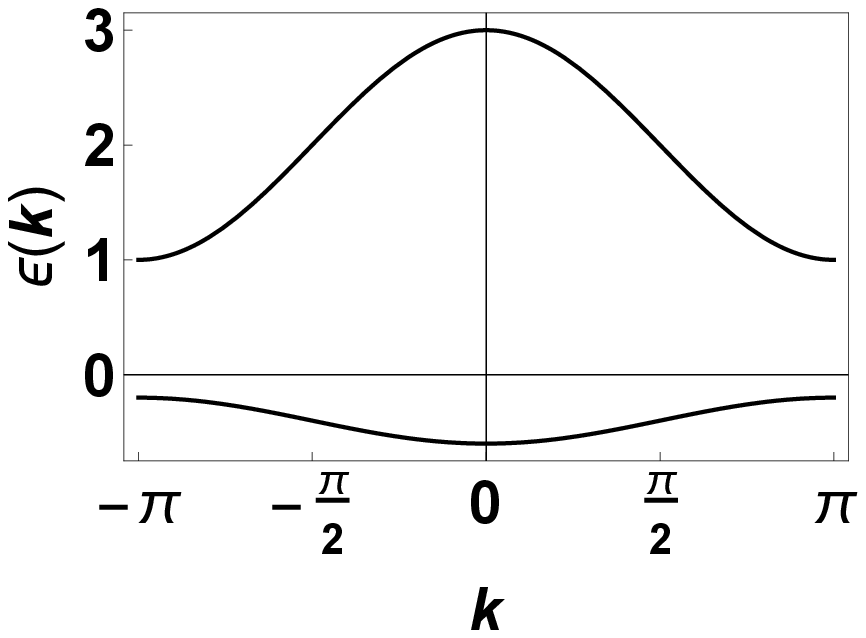}
		\end{minipage}
		
	\end{tabular}  
	\caption{\subref{fig:deltachain2} The lattice geometry of $H_{\mathrm{hop}}'$.
	The hopping amplitudes are given by $t_{1} = \nu(t+s)$, $t_{2} = \nu^{2} t$, and $t_{2}' = -\nu^{2} s$.
	Odd (black) and even (white) sites have on-site potentials $t-2\nu^{2}s$ and $-s + 2\nu^{2} t $, respectively.
	The corresponding energy bands are shown in \subref{fig:Energy band2} for $t = 1, \nu = 1/\sqrt{2}, s = 1/10$.
	}\label{fig:flatband2}
\end{figure}

%\subsection{Model 2}
%\label{ssec:model2}
%{\it Model 2.}---
So far, we have considered the flat-band model, but this is an idealized case in which the lowest energy band becomes completely dispersionless.
As a more realistic model, we consider a model with nearly flat band by adding a perturbation to the model in the previous section~\footnote{
It was proposed that the hopping part of the Hamiltonian with a nearly flat band can be realized with ultracold atoms in a sawtooth lattice~\cite{Zhang2015}.
}.
Here we define another Hubbard model on the same lattice as Theorem 1:
\begin{align}
H_{2} 
&= H_{\mathrm{hop}}' + H_{\mathrm{int}}, \label{ham2}
\end{align}
where $H_{\mathrm{hop}}'$ is defined as 
\begin{align}
H_{\mathrm{hop}}' 
&= -s \sum_{\alpha=1}^{n} \sum_{x \in \mathcal{E}} a_{x,\alpha}^{\dag} a_{x, \alpha}
+ t \sum_{\alpha=1}^{n} \sum_{x \in \mathcal{O}} b_{x,\alpha}^{\dag} b_{x, \alpha},  \label{hop2}
\end{align}
and $H_{\mathrm{int}}$ is defined in Eq.(\ref{hint}) with parameters $s, t, U > 0$.
When the hopping Hamiltonian $H_{\mathrm{hop}}'$ is written in terms of original fermion operator $c_{x, \alpha}$,  
it takes the form  $H_{\mathrm{hop}}' = \sum_{\alpha}\sum_{x,y \in \Lambda} t'_{xy} c_{x, \alpha}^{\dag} c_{y, \alpha}$
, where $t'_{2x-1, 2x-1} = t-2\nu^{2}s$, $t'_{2x,2x} = -s + 2\nu^{2}t$, $t'_{2x-1,2x} = t'_{2x,2x-1} = \nu(t+s)$, $t'_{2x-2, 2x} = t'_{2x, 2x-2} = \nu^{2}t$, $t'_{2x-1, 2x+1} = t'_{2x+1, 2x-1} = -\nu^{2}s$, and the remaining elements of $t'_{x, y}$ are zero.
When we consider the single particle problem, we obtain two bands with $\epsilon_{1}(k) = -s(2\nu^{2}+1) - 2\nu^{2}s \cos{k}$, $\epsilon_{2}(k) = t(2\nu^{2}+1) + 2\nu^{2}t \cos{k}$ (see Fig. \ref{fig:Energy band2}).
We see that the lowest band is no longer flat, however, it can be regarded as a nearly flat band when $s$ is small enough.
We focus on this model in the following and prove a theorem on the ferromagnetism.\par
%%%%%%%%%%%%%%%%%%%%%%%%%%%%%%%%%%%%%%%%%%%%%%%FIG%%%%%%%%%%%%%%%%%%%%%%%%%%%%%%%%%%%%%%%%%%%%%%%%%%%%%%
%%%%%%%%%%%%%%%%%%%%%%%%%%%%%%%%%%%%%%%%%%%%%%%%%%%%%%%%%%%%%%%%%%%%%%%%%%%%%%%%%%%%%%%%%%%%%%%%%%%%%%%%
%%%%%%%%%%%%%%%%%%%%%%%%%%%%%%%%%%%%%%%%%%%%%Theorem2%%%%%%%%%%%%%%%%%%%%%%%%%%%%%%%%%%%%%%%%%%%%%%%%%%
%%%%%%%%%%%%%%%%%%%%%%%%%%%%%%%%%%%%%%%%%%%%%%%%%%%%%%%%%%%%%%%%%%%%%%%%%%%%%%%%%%%%%%%%%%%%%%%%%%%%%%%%
%\subsection{Theorem 2.}
%\label{ssec:thm2}
{\it Theorem 2.}---Consider the Hamiltonian (\ref{ham2}) with the total fermion number $N_{\mathrm{f}} = M$.
For sufficiently large $t/s >0$ and $U/s > 0$, the ground states are the fully polarized states and unique apart from the trivial degeneracy due to the SU($n$) symmetry.

\smallskip

%:Proof of thm2
%\renewcommand{\proofname}{{\indent \it Proof of Theorem 2}}
%\renewcommand{\qedsymbol}{$\blacksquare$}
{\it Proof of Theorem 2.}---
First, we decompose the Hamiltonian (\ref{ham2}) into the sum of local Hamiltonians as 
\begin{align}
H_{2} = -sM(2\nu^{2}+1) + \lambda H_{\mathrm{flat}} + \sum_{x\in \mathcal{E}} h_{x}, \label{hamdecomposed}
\end{align}
where 
\begin{align}
H_{\mathrm{flat}} = \sum_{\alpha=1}^{n}\sum_{x \in \mathcal{O}} b_{x, \alpha}^{\dag} b_{x, \alpha} + \sum_{x \in \Lambda} \sum_{\alpha < \beta} n_{x,\alpha}n_{x, \beta}
\end{align}
and 
\begin{align}
&h_{x} \nonumber \\ 
&= \sum_{\alpha=1}^{n} \left(-s a_{x, \alpha}^{\dag} a_{x, \alpha} 
+ \frac{t-\lambda}{2} (b_{x-1, \alpha}^{\dag} b_{x-1, \alpha} + b_{x+1, \alpha}^{\dag} b_{x+1, \alpha})
\right) \nonumber \\
& + \frac{\kappa(U -\lambda)}{4}n_{x-2}(n_{x-2}-1) + \frac{U-\lambda}{4} n_{x-1}(n_{x-1}-1) \nonumber\\
& + \frac{(1-\kappa)(U-\lambda)}{2} n_{x}(n_{x}-1) + \frac{U-\lambda}{4} n_{x+1}(n_{x+1}-1) \nonumber\\
& + \frac{\kappa(U-\lambda)}{4}n_{x+2}(n_{x+2}-1) + s(2\nu^{2}+1), 
\label{local_ham}
\end{align}
where $n_{x}$ is defined as $n_{x} = \sum_{\alpha} n_{x, \alpha}$.
The two parameters $\lambda$ and $\kappa$ satisfy $0 < \lambda < \min\{t,U\}$ and $0 \leq \kappa < 1$.
To prove Theorem 2, we use the following lemmas.\par
%:Lemma1
{\it Lemma 1.}---Suppose the local Hamiltonian $h_{x}$ is positive semi-definite for any $x \in \mathcal{E}$. 
Then the ground states of the Hamiltonian (\ref{hamdecomposed}), and hence Eq. (\ref{ham2}),  are fully polarized states and unique apart from the trivial degeneracy due to the SU($n$) symmetry.\par
%Since the local Hamiltonian $h_{x}$ acts nontrivially only on a finite number of sites, one can check whether $h_{x}$ is positive semi-definite by numerically diagonalizing a finite dimensional matrix.
%The result for the SU(4) case is shown in Fig.\ref{fig:SU4boundary}.
%%%%%%%%%%%%%%%%%%%%%%%%%%%%%%%%%%%%%%%%%%%%%%%FIG%%%%%%%%%%%%%%%%%%%%%%%%%%%%%%%%%%%%%%%%%%%%%%%%%%%%%%
%\begin{figure}[H]
	%\centering
	%\includegraphics[width=0.6\columnwidth]{boundary_fig3.eps}
	%\caption{The positive semi-definiteness of $h_{x}$($n=4$) holds in the shaded region for $\nu=1/\sqrt{2}, \kappa = 0$.
	%The plot is obtained by diagonalizing $h_{x}$ numerically.
	%Lemma 1 says that the ground states of the full Hamiltonian $H_{2}$ are fully polarized states in the shaded region.}
	%\label{fig:SU4boundary}
%\end{figure}
%%%%%%%%%%%%%%%%%%%%%%%%%%%%%%%%%%%%%%%%%%%%%%%%%%%%%%%%%%%%%%%%%%%%%%%%%%%%%%%%%%%%%%%%%%%%%%%%%%%%%%%%
%%%%%%%%%%%%%%%%%%%%%%%%%%%%%%%%%%%%%%%%%%%%%%%%%%%%%%%%%%%%%%%%%%%%%%%%%%%%%%%%%%%%%%%%%%%%%%%%%%%%%%%%
%%%%%%%%%%%%%%%%%%%%%%%%%%%%%%%%%%%%%%%%%%%Summary%%%%%%%%%%%%%%%%%%%%%%%%%%%%%%%%%%%%%%%%%%%%%%%%%%%%%%
%%%%%%%%%%%%%%%%%%%%%%%%%%%%%%%%%%%%%%%%%%%%%%%%%%%%%%%%%%%%%%%%%%%%%%%%%%%%%%%%%%%%%%%%%%%%%%%%%%%%%%%%
%:Lemma2
{\it Lemma 2.}---Suppose that $t, U$ are infinitely large and $0 < \kappa < 1$.
Then the local Hamiltonian (\ref{local_ham}) is positive semi-definite.
(We take $\lambda$ and $\kappa$ to be proportional to $s$.)\par

We note that $h_{x}$ can be regarded as a finite dimensional matrix independent of the system size since the local Hamiltonian $h_{x}$ acts nontrivially only on a finite number of sites.
This means that the energy levels of $h_{x}$ depend continuously on the parameters.
Therefore, Lemma 2 guarantees that $h_{x}$ is positive semi-definite when $t, U$ are finite but sufficiently large.
%The energy levels of $h_{x}$ depend on parameters continuously because $h_{x}$ can be regard as a finite dimensional matrix.
%Therefore, Lemma 2 guarantees that $h_{x}$ is positive semi-definite when $t, U$ are finite but sufficiently large.
Then Lemma 1 implies that the ground states of the Hamiltonian (\ref{ham2}) are fully polarized states, which proves Theorem 2.
\hspace{\fill}$\blacksquare$
\par
Below, we prove Lemmas 1 and 2.
\par
%%%%%%%%%%%%%%%%%%%%%%%%%%%%%%%%%%%%%%%%%%%%%% proof of Lemma 1%%%%%%%%%%%%%%%%%%%%%%%%%%%%%%%%%%%%%%%%%%%%%%
%:Proof of Lemma 1
%\renewcommand{\proofname}{{\indent \it Proof of Lemma 1}}
%\renewcommand{\qedsymbol}{$\blacksquare$}
{\it Proof of Lemma 1.}---First, it is noted that a fully polarized state $\ket{\Phi_{\mathrm{all},1}} = \left(\prod_{x \in \mathcal{E}} a_{x, 1}^{\dag}\right) \ket{\Phi_{\mathrm{vac}}}$ satisfies $h_{x} \ket{\Phi_{\mathrm{all},1}} = 0$ for each $h_{x}$.
Since $h_{x}$ is SU($n$) invariant, all fully polarized states have zero energy.
We assume that $h_{x} \geq 0$ for all $x \in \mathcal{E}$. 
Let $\ket{\Phi_{\mathrm{GS}}^{\mathrm{flat}}}$ be an arbitrary ground state of $H_{\mathrm{flat}}$.
Since $H_{\mathrm{flat}} \ket{\Phi_{\mathrm{GS}}^{\mathrm{flat}}} = 0$ and $h_{x} \ket{\Phi_{\mathrm{GS}}^{\mathrm{flat}}} = 0$, we see that $H_{2} \ket{\Phi_{\mathrm{GS}}^{\mathrm{flat}}} = -s M(2\nu^{2} +1) \ket{\Phi_{\mathrm{GS}}^{\mathrm{flat}}}$.
From $h_{x} \geq 0$, the ground energy of $H_{2}$ is $-s M (2\nu^{2} +1)$.
If $\ket{\Phi_{\mathrm{GS}}}$ is an arbitrary ground state of $H_{2}$, it satisfies $H_{2} \ket{\Phi_{\mathrm{GS}}} = -s M(2\nu^{2} + 1)$.
From $H_{\mathrm{flat}} \geq 0$ and $h_{x} \geq 0$, we find $H_{\mathrm{flat}} \ket{\Phi_{\mathrm{GS}}} = 0$ and $h_{x} \ket{\Phi_{\mathrm{GS}}} = 0$.
This shows that any ground state of $H_{2}$ must be a ground state of $H_{\mathrm{flat}}$.
The Hamiltonian $H_{\mathrm{flat}}$ is nothing but the Hamiltonian $H_{1}$ with $t = U = 1$.
Thus, the ground states of $H_{2}$ are fully polarized and unique.
\hspace{\fill} $\blacksquare$
\par

\begin{figure}[H]
	\centering
	\includegraphics[width=0.6\columnwidth]{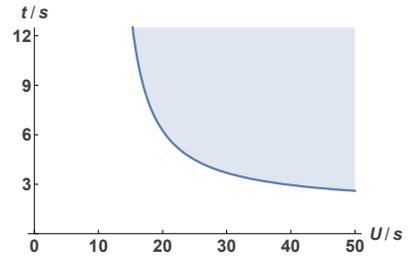}
	\caption{The positive semi-definiteness of $h_{x}$($n=4$) holds in the shaded region for $\nu=1/\sqrt{2}, \kappa = 0$.
	The plot is obtained by diagonalizing $h_{x}$ numerically.
	Lemma 1 says that the ground states of the full Hamiltonian $H_{2}$ are fully polarized states in the shaded region.
	For example, the ferromagnetism is established if $t/s \geq 4.5$ when $U/s = 25$.
	}
	\label{fig:SU4boundary}
\end{figure}

We remark that one can check whether $h_{x}$ is positive semi-definite by numerically diagonalizing a finite dimensional matrix.
%since the local Hamiltonian $h_{x}$ acts nontrivially only on a finite number of sites.
%Since the local Hamiltonian $h_{x}$ acts nontrivially only on a finite number of sites, one can check whether $h_{x}$ is positive semi-definite by numerically diagonalizing a finite dimensional matrix.
The result for the SU(4) case is shown in Fig.\ref{fig:SU4boundary}.
%%%%%%%%%%%%%%%%%%%%%%%%%%%%%%%%%%%%%%%%%%%%%%%FIG%%%%%%%%%%%%%%%%%%%%%%%%%%%%%%%%%%%%%%%%%%%%%%%%%%%%%%
%%%%%%%%%%%%%%%%%%%%%%%%%%%%%%%%%%%%%%%%%%%%%%%%%%%%%%%%%%%%%%%%%%%%%%%%%%%%%%%%%%%%%%%%%%%%%%%%%%%%%%%%
%%%%%%%%%%%%%%%%%%%%%%%%%%%%%%%%%%%%%%%%%%%%%%%%%%%%%%%%%%%%%%%%%%%%%%%%%%%%%%%%%%%%%%%%%%%%%%%%%%%%%%%%
%%%%%%%%%%%%%%%%%%%%%%%%%%%%%%%%%%%%%%%%%%%Summary%%%%%%%%%%%%%%%%%%%%%%%%%%%%%%%%%%%%%%%%%%%%%%%%%%%%%%
%%%%%%%%%%%%%%%%%%%%%%%%%%%%%%%%%%%%%%%%%%%%%%%%%%%%%%%%%%%%%%%%%%%%%%%%%%%%%%%%%%%%%%%%%%%%%%%%%%%%%%%%
%%%%%%%%%%%%%%%%%%%%%%%%%%%%%%%%%%%%%%%%%%%%%% proof of lemmma 2%%%%%%
%:Proof of Lemma 2
%\renewcommand{\proofname}{{\indent \it Proof of Lemma 2}}
%\renewcommand{\qedsymbol}{$\blacksquare$}
{\it Proof of Lemma 2.}---Due to the translational invariance, it suffices to show the case for $h_{0}$.
The local Hamiltonian $h_{0}$ is regarded as an operator defined on five sites $\{-2, -1, 0, 1, 2\}$, where $x = -2, -1$, and $0$ are identified with $x = 2M-2, 2M-1$, and $2M$, respectively.
On these sites, we define operators 
%$\tilde{a}_{-2,\alpha}:= \frac{1}{\sqrt{\nu^{2}+1}}(c_{-2, \alpha} - \nu c_{-1, \alpha})$, 
\begin{align}
\tilde{a}_{-2,\alpha}&:= \frac{1}{\sqrt{\nu^{2}+1}}(c_{-2, \alpha} - \nu c_{-1, \alpha}), \\
\tilde{b}_{-1, \alpha}&:= \nu c_{-2, \alpha} + c_{-1, \alpha} + \nu c_{0, \alpha}, \\ 
\tilde{a}_{0, \alpha}&:= -\nu c_{-1, \alpha} + c_{0, \alpha} + -\nu c_{1, \alpha}, \\ 
\tilde{b}_{1, \alpha}&:= \nu c_{0, \alpha} + c_{1, \alpha} + \nu c_{2, \alpha}, \\ 
\tilde{a}_{2, \alpha}&:= \frac{1}{\sqrt{\nu^{2}+1}}(-\nu c_{1, \alpha} + c_{2, \alpha}).
\end{align}
%$\tilde{a}_{-2, \alpha}:= (c_{-2, \alpha} - \nu c_{-1, \alpha})/\sqrt{\nu^{2}+1}$, 
%$\tilde{a}_{2, \alpha}:= \frac{1}{\sqrt{\nu^{2}+1}}(-\nu c_{1, \alpha} + c_{2, \alpha})$. 
%$\tilde{a}_{2, \alpha}:= (-\nu c_{1, \alpha} + c_{2, \alpha})/\sqrt{\nu^{2}+1}$.
These operators satisfy 
\begin{align}
\{\tilde{a}_{y, \alpha}, \tilde{a}_{y', \beta}^{\dag}\}
&= \begin{cases}
\delta_{\alpha,\beta} (2\nu^{2}+1) \ \ &\text{for}\ \  y=y'=0, \\
\delta_{\alpha,\beta} \frac{\nu^{2}}{\sqrt{\nu^{2}+1}}\ \  &\text{for}\ \ | y - y'| = 2, \\
\delta_{\alpha,\beta} \delta_{y,y'}\ \  &\text{for}\ \  y,y' = \pm2, 
\end{cases} \label{anticom2}\\ 
\{\tilde{a}_{y,\alpha}, \tilde{b}_{y', \alpha}^{\dag}\} &= 0.
\end{align}
Single-fermion states corresponding to these operators are linearly independent.
To show the lemma, we only need to consider states $\ket{\Phi}$ which have finite energy in this limit, i.e., $\lim_{t,U \rightarrow \infty} \bra{\Phi} h_{0} \ket{\Phi} < \infty$ .
The condition that $\ket{\Phi}$ has finite energy is equivalent to the following: 
\begin{align}
\tilde{b}_{y, \alpha} \ket{\Phi} &= 0 \ \ \text{for $y= \pm 1$}, \label{cond_b}\\
c_{y, \alpha} c_{y, \beta} \ket{\Phi} &= 0 \ \ \text{for $y = 0, \pm1, \pm2$}. \label{cond_c}
\end{align}
Let $\ket{\Phi}$ be a state which has finite energy.
From Eqs. (\ref{cond_b}) and (\ref{cond_c}) with $y=-2, 0, 2$, $\ket{\Phi}$ is written as 
\begin{align}
\ket{\Phi} = \! \! \! \sum_{\substack{A_{1} , \dots , A_{n} \subset \widetilde{\mathcal{E}}\\
A_{\alpha }\cap A_{\beta} = \emptyset
}} \! \! f(\{ A_{\alpha}\}) 
\! \left(\prod_{y \in A_{1}} \tilde{a}_{y,1}^{\dag}\right)
\! \dots \!
\left(\prod_{y \in A_{n}} \tilde{a}_{y,n}^{\dag}\right)
\! \ket{\Phi_{\mathrm{vac}}},
\end{align}
where $\widetilde{\mathcal{E}} = \{-2, 0, 2\}$ and $A_{\alpha}$ is an arbitrary subset of $\widetilde{\mathcal{E}}$.
Since $\widetilde{\mathcal{E}}$ contains three sites, the particle number of finite energy states must be less than or equal to three. 
Using the condition Eq. (\ref{cond_c}) with $y = \pm 1$, we see that all the finite energy states $\ket{\Phi}$ must be a fully polarized state 
%of 
over the five sites and have zero energy when the particle number is three.
For one-particle states, all the eigenvalues of $h_{0}$ are non negative.
Thus we only need to verify the positive semi-definiteness for two-particle sectors labeled by $(N_{\alpha}, N_{\beta}) = (2,0)$ and $(N_{\alpha}, N_{\beta}) = (1,1)$.
To this end, we solve the eigenvalue problem for $Ph_{0}P$ where $P$ denotes the projection operator onto the space of finite energy states.
In the sector $(2, 0)$, we find that there are three eigenstates 
\begin{align}
\ket{\Phi_{1}} &= \tilde{a}_{-2, \alpha}^{\dag} \tilde{a}_{0, \alpha}^{\dag} \ket{\Phi_{\mathrm{vac}}}, \\
\ket{\Phi_{2}} &= \tilde{a}_{0, \alpha}^{\dag} \tilde{a}_{2, \alpha}^{\dag} \ket{\Phi_{\mathrm{vac}}}, \\
\ket{\Phi_{3}} \! &= \! \left[
		\frac{\nu^{2}}{\nu^{2}+1} 
(\tilde{a}_{-2, \alpha}^{\dag}  \!-\! \tilde{a}_{2, \alpha}^{\dag}) \tilde{a}_{0, \alpha}^{\dag}
		\! - \! (2\nu^{2}+1) \tilde{a}_{-2, \alpha}^{\dag} \tilde{a}_{2, \alpha}^{\dag} 
		\right] \ket{\Phi_{\mathrm{vac}}}
\end{align}
%$\ket{\Phi_{3}} = \left(
%		\frac{\nu^{2}}{\nu^{2}+1} \tilde{a}_{-2, \alpha}^{\dag} \tilde{a}_{0, \alpha}^{\dag} 
%		\! + \frac{\nu^{2}}{\nu^{2}+1} \tilde{a}_{0, \alpha}^{\dag} \tilde{a}_{2, \alpha}^{\dag} 
%		\! - (2\nu^{2}+1) \tilde{a}_{-2, \alpha}^{\dag} \tilde{a}_{2, \alpha}^{\dag} 
%		\right) \ket{\Phi_{\mathrm{vac}}} $, 
%\begin{align}
%\ket{\Phi_{3}} \! = \! \left[
%		\frac{\nu^{2}}{\nu^{2}+1} 
%(\tilde{a}_{-2, \alpha}^{\dag}  \!-\! \tilde{a}_{2, \alpha}^{\dag}) \tilde{a}_{0, \alpha}^{\dag}
%		\! - \! (2\nu^{2}+1) \tilde{a}_{-2, \alpha}^{\dag} \tilde{a}_{2, \alpha}^{\dag} 
%		\right] \ket{\Phi_{\mathrm{vac}}}
%\nonumber
%\end{align}
and their corresponding eigenenergies are 0, 0, and $s(2\nu^{2} + 1)$, respectively.
In the sector $(1,1)$, there are four eigenstates.
Three of them can be obtained by applying $F^{\beta, \alpha}$ to the states $\ket{\Phi_{1}}, \ket{\Phi_{2}}$ and $\ket{\Phi_{3}}$.
As a state orthogonal to them, we get a singlet state 
\begin{align}
\ket{\Phi_{4}} = \left(\tilde{a}_{-2, \alpha}^{\dag}\tilde{a}_{2, \beta}^{\dag} - \tilde{a}_{-2, \beta}^{\dag} \tilde{a}_{2, \alpha}^{\dag} \right) \ket{\Phi_{\mathrm{vac}}},
\end{align}
and we find that this state satisfies 
\begin{align}
P h_{0} P \ket{\Phi_{4}} = s(2\nu^{2} + 1) \ket{\Phi_{4}}.
\end{align}
Clearly, the state $\ket{\Phi_{4}}$ has a positive energy.
Hence, we see that all the eigenvalues of $h_{0}$ are nonnegative.
Thus, we have proved Lemma 2.
\hspace{\fill} $\blacksquare$
\par
\section{Conclusion} \label{sec:conclusion}
%{\it Conclusion.}---
We have presented an extension of flat-band ferromagnetism to the SU($n$) Hubbard model on the railroad-trestle lattice.
%1D Tasaki lattice. 
Furthermore, we proved that in the nearly flat-band case, all the ground states are fully polarized if $t$ and $U$ are sufficiently large. 
%As a further extension, it will be possible to generalize our results into higher dimensional case. 
One can similarly construct and analyze models in higher dimensions, in which 
%Also in higher dimensional model, 
the ground states are fully polarized if the lowest band is completely flat. 
The previous results for the SU(2) Hubbard models in higher dimensions suggest that the parameter $\nu$ has to be larger than a threshold value $\nu_{c}>0$ when the lowest band is nearly flat~\cite{Shen1998, Tasaki2003}. The details will be discussed elsewhere.

%In this paper, we have studied the models whose energy gap is finite, whereas, it is possible to construct models with a gapless flat band.
%There is a possibility to extent the rigorous results for the flat-band ferromagnetism with a gapless band structure to the general SU($n$) case.
Although we have focused on models with a nonzero band gap, it would be interesting to see if the method developed in this paper can be extended to include SU($n$) Hubbard models with gapless flat or nearly flat bands~\cite{tanaka2003stability}. 
%\textbf{[H.K.: please cite here the paper by Tanaka-san about nearly flat band on kagome: Akinori Tanaka and Hiromitsu Ueda, Phys. Rev. Lett. \textbf{90}, 067204 (2003).] }
It would also be interesting to study SU($n$) ferromagnetism in systems with topological flat bands carryng nontrivial Chern number, as its SU($2$) counterpart has been discussed in~\cite{katsura2010ferromagnetism}. 
%In~\cite{katsura2010ferromagnetism}, the flat-band ferromagnetism with non-trivial Chern number is discussed.
%It would be intriguing to apply our discussion of the SU($n$) ferromagnetism to models with Chern bands.
Another direction for future research is to explore ferromagnetism in multiorbital Hubbard models, including the one with SU($n$) symmetry. In such systems, rigorous~\cite{li2014exact, li2015exact} and numerical results~\cite{xu2015sign} about ferromagnetism, which are different from the flat-band scenario, have been obtained recently. It is thus interesting to see to what extent our results can be generalized to the multiorbital case. 
%We have discussed single orbital Hubbard models.
%There are rigorous results about the ferromagnetism with multiorbital degrees of freedom ~\cite{li2014exact, li2015exact} and a related numerical study~\cite{ xu2015sign}, thus it is interesting to discuss the flat-band ferromagnetism in multiorbital SU($n$) Hubbard models.

%\medskip
\acknowledgments
We would like to thank Hal Tasaki and Akinori Tanaka for valuable discussions. H.K. was supported in part by JSPS Grant-in-Aid for Scientific Research on Innovative Areas: No. JP18H04478 and JSPS KAKENHI Grant No. JP18K03445.
%{\it Acknowledgements.}---
%%%%%%%%%%%%%%%%%%%%%%%%%%%%%%%%%%%%%%%%%%%%%%%%%%%%%%%%%%%%%%%%%%%%%%%%%%%%%%%%%%%%%%%%%%%%%%%%%%%%%%%%
%:bibliography
%\begin{thebibliography}{1}
%\bibitem{example} authors, Phys. Rev. B {\bf 78}, 195125 (2008).
%\end{thebibliography}
\bibliographystyle{apsrev4-1}
\bibliography{reference}
%%%%%%%%%%%%%%%%%%%%%%%%%%%%%%%%%%%%%%%%%%%%%%%%%%%%%%%%%%%%%%%%%%%%%%%%%%%%%%%%%%%%%%%%%%%%%%%%%%%%%%%%
%\begin{widetext}
%Supplementary Materials
%\begin{center}
%\Large{Supplemental Material for: {\it Supplemental Material} }
%\end{center}

%\begin{center}
%\bf{Supplemental Material 1}
%\end{center}
%Main contents of Supplemental Material 1

%\begin{center}
%\bf{Supplemental Material 2}
%\end{center}
%Main contents of Supplemental Material 2

%\begin{center}
%\bf{Supplemental Material 3}
%\end{center}
%Main contents of Supplemental Material 3
%\end{widetext}

\end{document}